\documentclass[twocolumn,trackchanges]{aastex61}

\def\Ha{H$\alpha$}

\def\hi{{\sc H\thinspace i}}

\def\msol{\rm\,M_\odot}

\def\kms{{\rm\,km\,s^{-1}}}

\def\vtot{v_\perp}
\def\vloctot{v_{\rm loc,\perp}}

\accepted{October 13, 2018}
\submitjournal{ApJL}

\shorttitle{OB Kinematics and Runaways the SMC}
\shortauthors{Oey et al.}

\begin{document}

\title{Resolved Kinematics of Runaway and Field OB Stars in the Small Magellanic Cloud}


\correspondingauthor{M. S. Oey}
\email{msoey@umich.edu}

\author[0000-0002-5808-1320]{M. S. Oey}
\affil{Department of Astronomy \\
University of Michigan \\
1085 South University Ave. \\
Ann Arbor, MI   48109-1107, USA}

\author{J. Dorigo Jones}
\affil{Department of Astronomy \\
University of Michigan \\
1085 South University Ave. \\
Ann Arbor, MI   48109-1107, USA}

\author{N. Castro}
\affil{Department of Astronomy \\
University of Michigan \\
1085 South University Ave. \\
Ann Arbor, MI   48109-1107, USA}
\affil{Present address: \\
  Leibniz-Institut f\"ur Astrophysik \\
  An der Sternwarte 16 \\
  14482 Potsdam, Germany}

\author{P. Zivick}
\affil{Department of Astronomy \\
University of Virginia \\
530 McCormick Rd. \\
Charlottesville, VA   22904, USA}

\author{G. Besla}
\affil{Steward Observatory \\
University of Arizona \\
933 N. Cherry Ave., \\
Tucson, AZ   85721, USA}

\author{H. C. Januszewski}
\affil{Department of Astronomy \\
University of Michigan \\
1085 South University Ave. \\
Ann Arbor, MI   48109-1107, USA}
\affil{Present address:  \\
  Gemini Observatory \\
  670 N. A'ohoku Place, \\
  Hilo, HI 96720, USA}

\author{M. Moe}
\affil{Steward Observatory \\
University of Arizona \\
933 N. Cherry Ave., \\
Tucson, AZ   85721, USA}

\author{N. Kallivayalil}
\affil{Department of Astronomy \\
University of Virginia \\
530 McCormick Rd. \\
Charlottesville, VA   22904, USA}

\author{D. J. Lennon}
\affil{ESA, European Space Astronomy Centre \\
  Apdo. de Correos 78\\
  E-28691 Villanueva de la Ca\~nada, Madrid,
  Spain}
\affil{Instituto de Astrof\'isica de Canarias\\
  E-38205 La Laguna, Tenerife, Spain}

\begin{abstract}
We use {\sl GAIA} DR2 proper motions of the RIOTS4 field OB stars
in the Small Magellanic Cloud (SMC) to study the kinematics of runaway stars.  The data reveal
that the SMC Wing has a systemic peculiar motion relative to the SMC Bar of
$(v_\alpha, v_\delta)=(62\pm 7, -18\pm5)\ \kms$
and relative radial velocity $+4.5\pm 5.0\ \kms$.  This unambiguously
demonstrates that these two regions are kinematically distinct: {\it the
Wing is moving away from the Bar,} and towards the Large Magellanic
Cloud with a 3-D velocity of $64\pm10\ \kms$.
This is consistent with models for a recent, direct collision between the Clouds.
We present transverse velocity distributions for our field OB stars, confirming
that unbound runaways comprise on the order of half our sample, possibly more.
Using eclipsing binaries and double-lined
spectroscopic binaries as tracers of dynamically ejected runaways, and
high-mass X-ray binaries (HMXBs) as tracers of runaways accelerated by supernova
kicks, we find significant contributions from both populations.
The data suggest that HMXBs have lower velocity dispersion relative to
dynamically ejected binaries, consistent with the former corresponding
to less energetic supernova kicks that failed to unbind the components.
Evidence suggests that our fast runaways are dominated by dynamical,
  rather than supernova, ejections.
\end{abstract}



\section{Introduction:  Runaway OB Stars} \label{sec:intro}

Field OB stars constitute a significant subset of the massive star
population in star-forming galaxies.  Given the power-law cluster mass
distribution, \citet{oey2004} showed that field OB stars typically
comprise 20 -- 30\% of massive stars.  However, the field additionally
includes significant numbers of high-velocity, runaway stars
ejected from clusters \citep{blaauw1961,hoogerwerf2000}.  The classic work by
\citet{blaauw1961} found that about 20\% of early B stars and 3\% of O
stars are runaways, and \citet{Moffat1998} find a runaway
fraction of O and Wolf-Rayet stars of 14\% from {\sl 
HIPPARCOS} space velocities.  However, infrared work by
\cite{deWit2005} suggests that over 90\% of O stars are runaways,
and some studies suggest that {\it all} truly
isolated field massive stars are runaways
\citep[e.g.,][]{Pflamm-Altenburg2010,
gvaramadze2011b}.  On the other hand, a variety of observational
evidence suggests that field objects formed in relative isolation are
also a major, if not dominant, component of the field massive star 
population \citep[e.g.,][]{Lamb2016,Oey2013}.

Two principal mechanisms are responsible for generating runaway
stars.  One is dynamical ejection from gravitationally unstable
configurations \citep{Poveda1967, Leonard1988};
another is the acceleration of a star when its binary
companion explodes as a supernova \citep[SN;][]{blaauw1961}.
These are dominated, for higher velocity runaways, by explosions
  that generate a recoil ``kick'' to the companion, rather than simple
  ``slingshot'' acceleration \citep[e.g.,][]{Renzo2018}.  
The relative importance of the dynamical vs SN mechanisms is poorly
known.  For the latter, a minority of runaways should retain
their neutron star companions, while SNe disrupt most of these binaries
\citep[e.g.,][]{Brandt1995,Renzo2018}.  This is supported by searches
for runaways with neutron star companions \citep[e.g.,][]{Philp1996, Sayer1996}. 

The dynamical ejection mechanism takes place 
primarily via binary-binary interactions
\citep{Poveda1967, Leonard1988}. 
This is the only process that can yield binary runaways consisting of
two non-compact stars, in addition to single runaways.
From 42 runaways in the Galactic field O star sample, 
the frequency of non-compact, multiple runaways is at least $\sim15$\% of
O star runaways, based on detections of double-lined spectroscopic binaries (SB2) and
astrometric binaries \citep{Mason2009}.  If such a value is confirmed, 
dynamical ejection may well dominate the massive runaway population.
However, dynamical ejection may require unusual mass ratios and orbital
parameters \citep{Leonard1990}.  For example, for a binary-binary
ejection model, \citet{Clarke1992} require nearly all O stars
to have close binary companions with mass ratios $>0.25$, to achieve
a runaway fraction $>10$\%.  Although most O stars indeed have
close OB companions \citep{Sana2012}, there is a
significant contribution from lower-mass close companions as well \citep{Moe2015}. 

Hence, the statistical properties and fundamental parameters of the OB 
runaway population offer critical diagnostics of the ejection
mechanisms, and their statistics 
also depend strongly on cluster properties and dynamical evolution
\citep[e.g.,][]{Poveda1967, Hills1980}.
Evaluating the frequency and properties of runaways is therefore vital to
understanding the nature of both the field population and clusters
\citep[e.g.,][]{Clarke1992, PortegiesZwart2000}.
However, testing such predictions has been limited to date by 
the substantial uncertainties for runaway statistics and
inhomogeneous data in the Milky Way.

\begin{figure*}
\epsscale{1.1}
\plotone{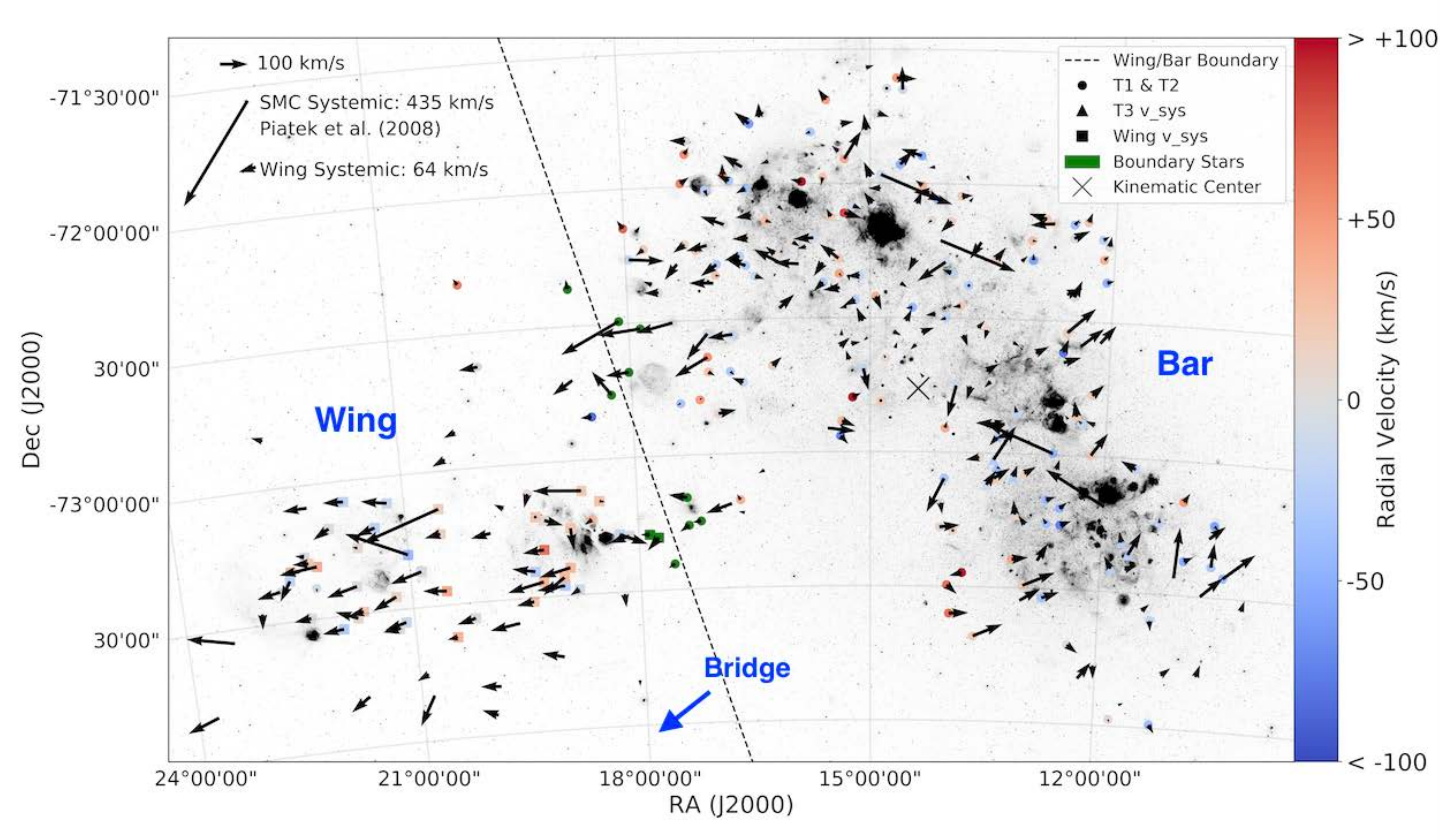}
\plotone{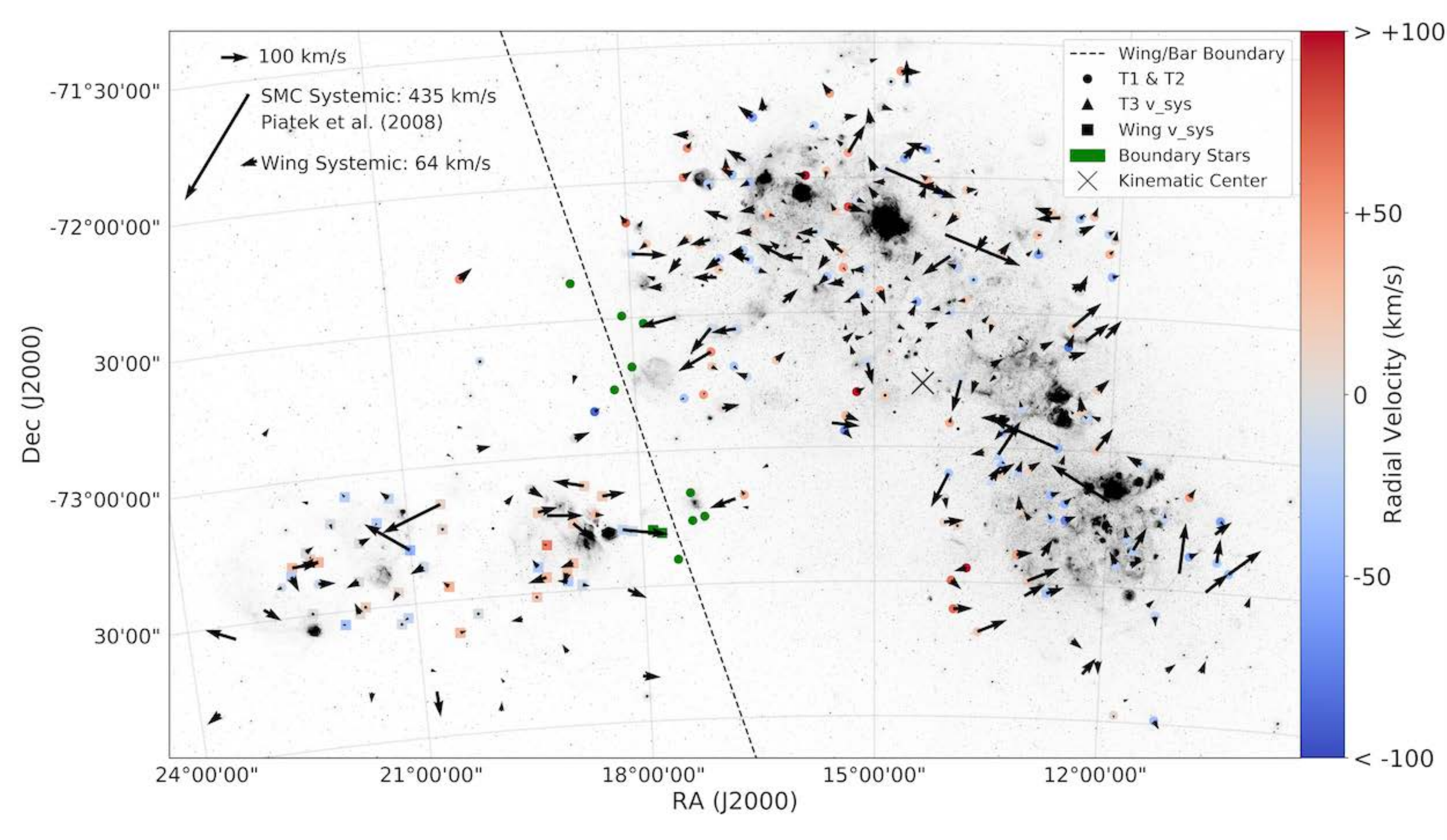}
\caption{
Vector map of 315 {\sl GAIA} DR2 PM residuals for RIOTS4 field OB stars
superposed on \Ha\ image from \citet{Smith2005}.
Panel ($a$, top) shows only corrections for geometric perspective and SMC
systemic velocity.  Panel ($b$, bottom) additionally includes a separate
peculiar PM and RV correction for Wing stars.  The adopted boundary between the
so-called Wing and Bar regions is shown by the dashed line, with stars removed from the
PM samples indicated in green.  The vectors for the Wing peculiar
PM correction and SMC systemic PM are shown, as is the adopted
kinematic center.  Colors show available RVs for 216 stars.  Symbols
indicate RV source, with triangles and squares showing 
stars with systemic RV measurements from multi-epoch monitoring; 
T1, T2 and T3 refer to data from tables in \citet[][see text, Section 2]{Lamb2016}.
The direction toward the Magellanic Bridge is indicated.
\label{f_vectors}}
\end{figure*}

Here, we examine the kinematics of
runaway OB stars in an extragalactic environment:  the Small
Magellanic Cloud (SMC), where statistical completeness is easily
evaluated.  {\sl GAIA} DR2 \citep{lindegren2018} now offers an outstanding
data set of proper motions for the SMC, which
is located at high Galactic latitude and low extinction, owing to its
low metallicity.  Our study is based on the Runaways and Isolated
O-Type Star Spectroscopic Survey of the SMC (RIOTS4)
which is yielding a detailed, quantitative characterization of this
field OB population \citep[e.g.,][]{Lamb2016}.
This sample also offers an opportunity to look at 
  large-scale stellar kinematics of the SMC's young population.

\section{RIOTS4 Proper Motions from {\sl GAIA}}

The RIOTS4 field star sample is defined from \citet{oey2004}, who
used the $UBVR$ photometric survey of the SMC by \citet{Massey2002} to
identify OB-star candidates based on having reddening-free parameter
$Q_{UBR}\leq -0.84$ and $B\leq 15.21$.  These serve as uniform
selection criteria for stars earlier than spectral type $\sim$B0.5.
Field and cluster stars were defined using the friends-of-friends
algorithm of Battinelli (1991), adopting a clustering length of 28 pc,
which yields 374 field stars, or 28\% of all SMC OB stars identified by
\citet{oey2004}.  An additional 23 O stars in the RIOTS4 sample were  
identified using UV photometric criteria on data from the {\sl Ultraviolet Imaging
  Telescope}, yielding a total of 397 stars.  The two subsamples are
given in Tables 1 and 2, respectively, of \citet[][T1, T2]{Lamb2016}.
Spectroscopic observations of these RIOTS4 stars were obtained at Magellan
using the IMACS and MIKE spectrographs  \citep{Lamb2016}, with
on-going, multi-year monitoring in the Wing region using the M2FS multi-fiber
spectrograph.  The latter data yield systemic radial velocities (RVs) for
detected binary systems; a smaller region in the SMC Bar was similarly
monitored with IMACS, and systemic RVs reported in Table~3 of
\citet[][T3]{Lamb2016}.  

We identify the RIOTS4 stars in the {\sl GAIA} DR2 
catalog by specifying a position match within $3\arcsec$ and
magnitude match $|G-V|<0.3$,
yielding 328 matches.
We further vet the sample by including only
stars having both RA and Dec proper motion (PM) errors $<1\sigma$ from
the median error,
eliminating 12 stars.  We also delete one more star that has 
  RA or Dec errors $>3.5 \sigma$ after the initial clip.
The final RA and Dec standard
deviations are 55 $\kms$ and 37 $\kms$, respectively.
Our stars are generally in the range $12.5 < G < 15.5$.
Figure~\ref{f_vectors}$a$ depicts the PMs for these 315
RIOTS4 stars, adopting a mean SMC distance modulus of 18.99 \citep{cioni2000}
(Table~\ref{t_master}).
The PMs are residuals relative to the SMC 
systemic PM of $(\mu_\alpha,\mu_\delta)=(0.754,-1.252)$ mas yr$^{-1}$
from  \citet{Piatek2008}, which is the published value 
that minimizes residuals in the SMC Bar.  The shown vectors are also
corrected for geometric perspective using the model of
\citet{vandermarel2002}, but adopting a center of motion at the midpoint
between the \hi\ \citep{stanimirovic2004} and stellar 
\citep{ripepi2017} kinematic centers.  Figure~\ref{f_vectors} reveals
a pattern of motion consistent with that of \citet{zivick2018}.
We also apply geometric corrections to the RVs and subtract
the median SMC systemic RV of 152 $\kms$ for the Bar.

It is apparent in Figure~\ref{f_vectors} that the SMC Wing shows a median systemic PM of
$(\mu_\alpha, \mu_\delta) = (0.207\pm 0.025, -0.060\pm 0.016)\ \rm mas\ yr^{-1}$,
corresponding to velocities
$(v_\alpha, v_\delta)=(62\pm7, -18\pm5)\ \kms$ and 
a total transverse velocity $\vtot=64\pm 8 \kms$.  This Wing peculiar
motion is obtained after deleting 11 stars within 0.5 degree of the
boundary shown in Figure~\ref{f_vectors}.  Panel $b$ 
shows the PMs with the 68 Wing stars corrected for this additional
peculiar motion.  This effect is robust to the choice of systemic PM and 
kinematic center.
In applying geometric corrections to the RVs, we find that the RV
  offset reported by \citet{Lamb2016}, who did not correct for
  perspective, is due primarily to this effect.  The Wing median RV offset
  from our data is now $+4.5\pm 5.0\ \kms$.

\begin{deluxetable*}{cccccccccccccc}
\tablecaption{Kinematic Data for RIOTS4 Field OB
  Stars\tablenotemark{a} \label{t_master}}
\tablewidth{0pt}
\tablehead{
  \colhead{ID\tablenotemark{b}} & \colhead{Subgroup\tablenotemark{c}} &
  \colhead{$\vtot$\tablenotemark{d}} &
  \colhead{$\vloctot$\tablenotemark{d}} & \colhead{Quality\tablenotemark{e}} &
  \colhead{RV\tablenotemark{d}} &
  \colhead{$v_{\rm RA}$\tablenotemark{f}} &
  \colhead{err\tablenotemark{g}} & \colhead{$v_{\rm Dec}$\tablenotemark{f}} &
  \colhead{err\tablenotemark{g}} & \colhead{$v_{\rm loc,\rm RA}$\tablenotemark{f}} &
  \colhead {err\tablenotemark{g}} &
  \colhead{$v_{\rm loc,\rm Dec}$\tablenotemark{f}} & \colhead{err\tablenotemark{g}}\\
  \colhead{- } & \colhead{- } & \colhead{$\kms$} &
  \colhead{$\kms$} & \colhead{-} &
  \colhead{$\kms$} &  \colhead{$\kms$} &
  \colhead{$\kms$} & \colhead{$\kms$} &
  \colhead{$\kms$} & \colhead{$\kms$} &
  \colhead{$\kms$} & \colhead{$\kms$} &
  \colhead{$\kms$}} 
 \startdata
107 & -,-,-,B,- & 42 & 20 & -,0,0 & ... & 186 & 17 & -347 & 12 & 167 & 5 & -349 & 7 \\
1037 & -,-,-,B,- & 145 & 99 & -,0,0 & -43 & 85 & 24 & -300 & 18 & 166 & 4 & -356 & 4 \\
1600 & -,E,-,B,- & 22 & 43 & -,0,0 & -58 & 188 & 22 & -401 & 21 & 187 & 5 & -358 & 4 \\
1631 & -,-,-,B,- & 87 & 51 & -,0,0 & -33 & 190 & 17 & -298 & 15 & 175 & 2 & -347 & 7 \\
1830 & -,-,-,B,- & 75 & 32 & -,0,0 & ... & 143 & 20 & -341 & 17 & 174 & 4 & -349 & 6 \\
2034 & -,-,-,B,- & 47 & 21 & -,0,0 & ... & 186 & 21 & -339 & 15 & 174 & 2 & -357 & 5 \\
2093 & -,-,-,B,- & 105 & 63 & -,0,0 & ... & 122 & 20 & -318 & 18 & 178 & 5 & -347 & 8 \\
3224 & -,-,-,B,- & 39 & 9 & -,0,0 & -51 & 169 & 15 & -368 & 15 & 169 & 3 & -359 & 4 \\
3459 & -,-,-,B,- & 39 & 7 & -,0,0 & 50 & 183 & 21 & -353 & 17 & 179 & 5 & -359 & 3 \\
3815 & -,-,-,B,- & 181 & 147 & -,0,1 & ... & 182 & 21 & -204 & 20 & 181 & 4 & -351 & 3 \\
\enddata
\tablenotetext{a}{Table~\ref{t_master} is available in its entirety on-line.}
\tablenotetext{b}{From \citet{Massey2002}}
\tablenotetext{c}{`E', `S', `X' indicate EB, SB2 and HMXB,
respectively; `B', `W', `D' indicate Bar, Wing, and boundary stars,
respectively; `m' indicates object in multi-epoch spectroscopic sample.}
\tablenotetext{d}{
Final residual velocity relative to SMC and Wing systemic motion.
  The $\vloctot$ values are computed relative to local velocity fields
  (see text).  RV errors are typically $10\ \kms$.}
\tablenotetext{e}{`a' indicates object meets $\sim 10$\% asymmetry
  criterion in RA vs Dec; the second and third values give the number
  of stars within 1$\arcsec$ and 1.5$\arcsec$, respectively.}
\tablenotetext{f}{Transverse velocity computed from proper
    motion, without geometric or systemic velocity corrections.
\tablenotetext{g}{Measurement errors, not including systematic errors
  (see text).}}
\end{deluxetable*}

Table~\ref{t_master} lists the total residual transverse velocities
$\vtot$ and RV, along with non-residual PMs for our field OB stars.
We also give locally determined transverse velocities (see \S 3.2).
Stars in the Wing region (Figure~\ref{f_vectors}) are indicated, and
their residual values are corrected for 
the Wing peculiar motions in PM and RV.  All PM values are based
on the original SMC geometric correction described above; our models
show that a specific correction for the Wing would modify the
velocities by at most 3.5 $\kms$, whereas systematic errors on the
geometric correction are on the order of 30 $\kms$. 
The FWHM of the Cepheid distance distribution yields a variation of 15\%
\citep[e.g.,][]{ripepi2017}, with extremes up to 50\%,
given the SMC's end-on orientation to the line of sight.  These
imply distance uncertainties that propagate directly to our transverse
velocities.

\section{SMC field OB kinematics}

\subsection{Proper motion of the SMC Wing}

While it is necessary to correct for the Wing's systemic motion to 
  identify runaway stars,
our data also clearly reveal that {\it the Wing and Bar are
  kinematically distinct components,} with a 3-D offset of
$64\pm10\ \kms$.
The Wing has been identified as the southeast component of the
  SMC, and extends $\gtrsim 2^\circ$ beyond our observed data set,
  merging with the Magellanic Bridge linking the SMC to the Large
  Magellanic Cloud (LMC).
  Previous work \citep[e.g.,][]{Brueck1978, Dobbie2014} shows 
an older coexisting Wing stellar population having RVs similar to those of
 young stars.  This suggests  that our sample is a good tracer of the bulk motion of
 this region, but follow-up examination of {\sl GAIA} PMs for red giant
 stars is needed to understand differentials between the old and young populations.

The dynamical state of the Wing
provides a vital kinematic discriminant for dynamical models of the internal
structure of the SMC, the recent encounter history of the Magellanic
Clouds and formation of the Magellanic Bridge
\citep[e.g.,][]{Besla2012,zivick2018}.
In particular, the Wing kinematics seen here are consistent with  
transverse motion along the Bridge towards the LMC, instead of
perpendicular to the Bridge.  This 
vividly confirms models for a recent,
direct collision between the Clouds
100 -- 200 Myr ago,
for which gas velocities are
expected to be aligned with the Bridge.  In contrast, motions
perpendicular to the Bridge are theoretically expected in a tidal
stripping scenario of an SMC that did not collide with the LMC,
allowing it to retain ordered rotation.  The absence of perpendicular
motion is consistent with the results of \citet{zivick2018}, who find
little evidence of rotation in the SMC,
which also supports the direct collision model.

\subsection{Field OB kinematics and runaway stars}

\begin{figure*}
\epsscale{1.2}
\plotone{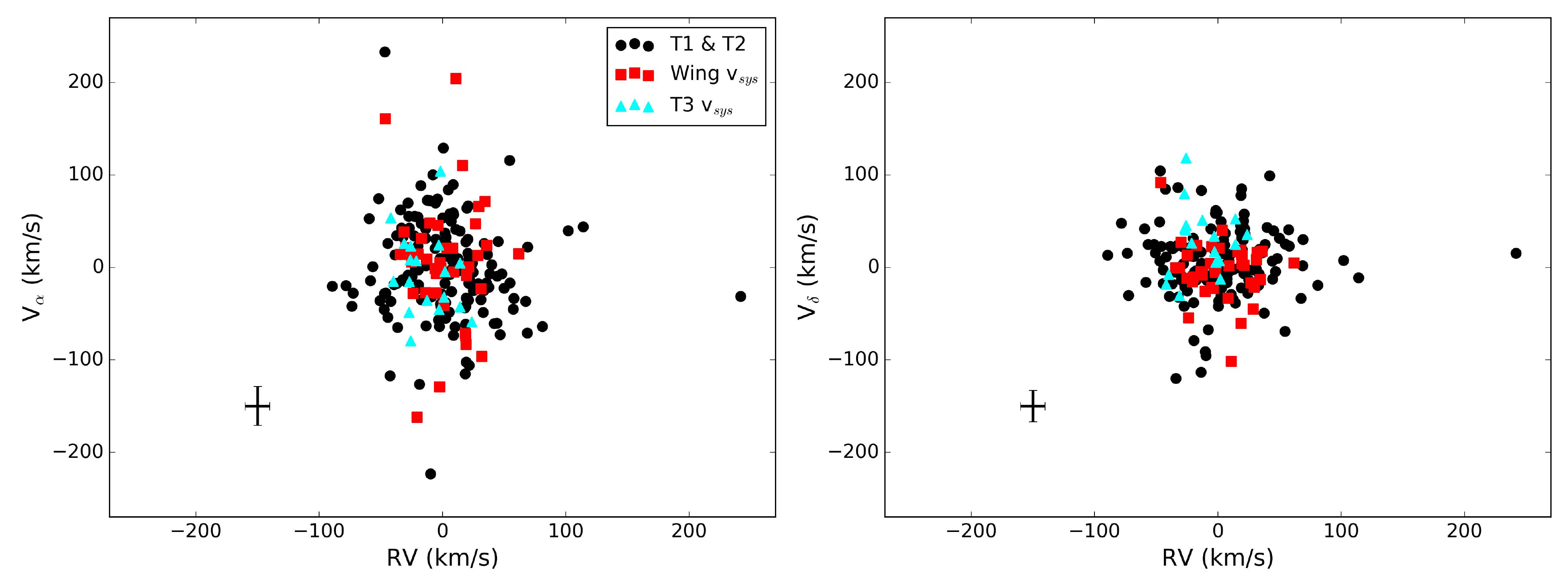}
\caption{
Comparison of residual RVs with ($a$) RA and ($b$) Dec velocities.  Triangles and squares
show objects with systemic RVs obtained through multi-epoch monitoring.
The remainder of the sample are single-epoch RV measurements.
Subsample labels are as in Figure~\ref{f_vectors}.
\label{f_pmrv}}
\end{figure*}

\begin{figure*}
\epsscale{1.2}
\plotone{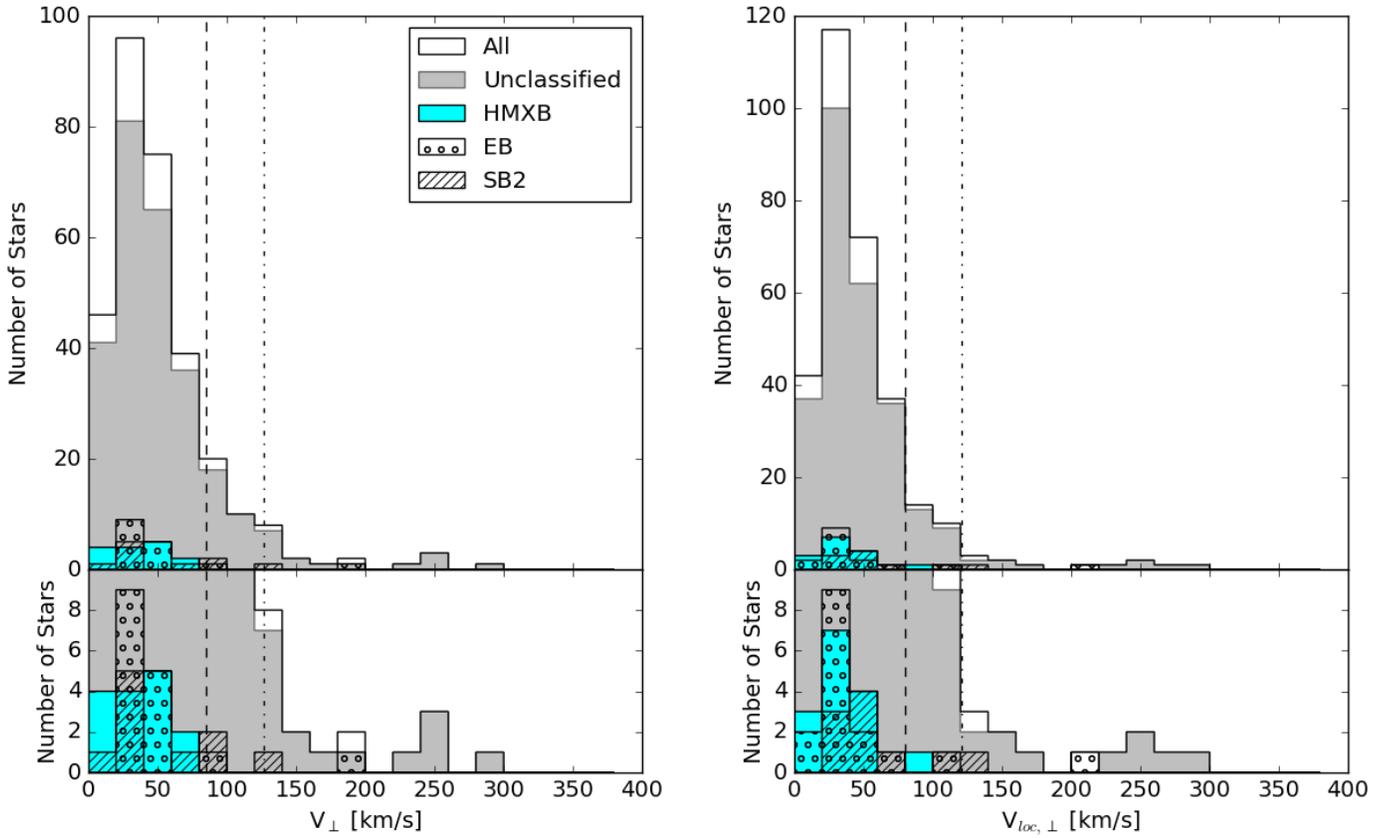}
\caption{
  Distribution of ($a$) $\vtot$ and ($b$) $\vloctot$.
  EB, SB2, and HMXB populations are shown as
  indicated.  The vertical lines correspond to 1-$\sigma$ and
  2-$\sigma$ velocities from the medians, using values in
  Table~\ref{t_binaries}.  The lower insets show the $y$-axis
  zoomed for clarity.
  \label{f_histkine}}
\end{figure*}

As noted in \S 1, non-compact binaries that are runaway systems must
result from the dynamical ejection mechanism.
In addition to SB2s, non-compact binaries also can be 
identified as eclipsing binaries (EBs), which are given
in the OGLE-III EB catalog for the SMC
\citep{Pawlak2016}.  Of the 315 stars, only 295 are covered by the
OGLE-III survey, since the easternmost Wing region is excluded.
Furthermore, it is also possible to identify
binaries with a compact remnant as high-mass X-ray binaries (HMXBs),
which are compiled by \citet{Haberl2016}.  Runaway HMXBs result from the
SN acceleration mechanism.  Stars found to be SB2, EB, and/or
HMXB are indicated in Table~\ref{t_master}.

In Table~\ref{t_binaries}, we list the numbers of stars in each
population, along with the mean and median $\vtot$ and
standard deviations $\sigma$ for each binary population.
The median {\sl GAIA} errors in
$v_\alpha,\ v_\delta,$ and $\vtot$
for the sample are 22, 17, and 28 $\kms$, with 
standard deviations on the errors of 5, 3, and 6 $\kms$, 
respectively; thus the errors are more than a factor of 2 below the
respective observed $\sigma_\alpha,\ \sigma_\delta,$ and $\sigma_{\perp}$.
Therefore, these standard deviations reflect actual velocity dispersions
convolved with the substantial errors.
Figure~\ref{f_pmrv} compares PM and RV for the 207
stars that also have RVs reported by \citet{Lamb2016}, omitting the stars
near the Wing-Bridge boundary.  Stars
shown by triangles and squares are, respectively, systemic RVs estimated from our 
multi-epoch monitoring surveys of the Bar
(Table~3 of \citet{Lamb2016})
and Wing region (on-going).  We caution that the single-epoch
observations often include significant binary motions.
Figure~\ref{f_pmrv} confirms that the measured PMs and RVs are comparable.
It is apparent that the PMs show a larger spread in RA than Dec, which
is due to the asymmetric {\sl GAIA} errors \citep[e.g.,][]{lindegren2018}.
Thus, velocities, especially large ones, for any given star may not be
real, but the kinematics of subsamples may be compared.  We have also
inspected the OGLE-III \citep{udalski2008} images
available for 304 of our target stars  to evaluate PSF asymmetry and
crowding, which can degrade the astrometry; stars that may be thus affected
are flagged in Table~\ref{t_master}.  Measured $\vtot$ have no
apparent dependence on PSF asymmetry when flagging those with $\gtrsim 10$\%
variation in RA vs Dec.  But whereas $\sim$10\% of all targets
have neighbors within 1.5$\arcsec$, targets having $\vtot >
1\sigma_\perp$ are much more likely (24\%, 10 out of 41) to have such
close neighbors.  Thus we caution that stars
with $\vtot\gtrsim 200\ \kms$ are likely dominated by spurious values
\citep[e.g.,][]{Platais2018}.  

Figure~\ref{f_histkine}$a$ shows $\vtot$
distributions for the 304 stars, which exclude those
near the Wing-Bar boundary.  We also show the
contribution of each binary population.  The peak
of the $\vtot$ distribution
occurs at the value corresponding to the median error of 28 $\kms$, 
reflecting the {\sl GAIA} {\bf detection} limit at $G<15$.
The median $\vtot$ of 44 $\kms$ is therefore significant, occurring in
the runaway velocity regime, since the typical velocity dispersion in OB
associations is $\sim5\ \kms$ \citep[e.g.,][]{Melnik2017}.
Since half the sample has $\vtot$ greater than the median value, 
  which in turn is larger than typical bound velocity
  dispersions, this therefore implies that well over half the sample
  corresponds to unbound runaways, since many unbound stars also occur
  at velocities below the median \citep[e.g.,][]{Renzo2018}.

The population of ``unclassified'' stars
in Table~\ref{t_binaries} simply refers to
the remainder of the sample excluded by the other categories, and
therefore includes any unidentified binaries.  Thus, the EB, SB2, and HMXB
populations are lower limits on the true numbers of non-compact and
compact binary systems.  Table~\ref{t_binaries} identifies 22
non-compact systems and 15 compact.
We caution that the field includes a likely
substantial population of non-runaway stars that formed in situ
\citep[e.g.,][]{Oey2013}; analysis of the binary
frequencies will be presented in a future work.
Figure~\ref{f_histkine} and Table~\ref{t_binaries} show that the
kinematics of the EB and SB2 populations are generally consistent with
those of the total population, showing similar transverse velocity
dispersions.  In contrast, the spreads
for the HMXBs are much lower, with the non-compact binaries
having values about 50\% larger than for the HMXBs.  In fact,
none of the HMXBs have $\vtot >1\sigma$ from the median of the total sample
(Figure~\ref{f_histkine}$a$).  We caution that one SB2 with velocity
$> 1\sigma$, star 76253, has another star within $1.5\arcsec$ to the
north, which may affect the {\sl GAIA} astrometry (Table~\ref{t_master}).

The above kinematics are derived from only two assumed systemic
components, Wing and Bar, as described in \S 2.  Since there may be additional,
higher-order systemic motions, we also examine the PMs of our
sample stars relative to their local velocity fields.  We use
the {\sl GAIA} PMs of stars from the \citet{Massey2002} catalog of OB
stars within a 5$\arcmin$ (90 pc) radius of the target 
star to obtain the mean local velocity of the young
population.  We fitted the local PM distributions in RA and Dec
with gaussians having $\sigma=45\ \kms$ and 55 $\kms$,
respectively; these are the mean values for the Bar.
The local transverse velocities $v_{\rm loc}$ obtained in this
  way are given in Table~\ref{t_master}.
Figure~\ref{f_histkine}$b$ and Table~\ref{t_binaries} show the
resulting residual PM kinematics.
We see a similar pattern as before, with the HMXBs again showing smaller
standard deviations $\sigma_{\rm loc}$ when measured relative to the
local fields.

\begin{deluxetable*}{lccccr}
\tablecaption{Kinematics of Binary SMC Field OB Stars \label{t_binaries}}
\tablewidth{0pt}
\tablehead{
  \colhead{\ } &
  \colhead{Unclassified\tablenotemark{a}} &
  \colhead{EB} &
  \colhead{SB2} & 
  \colhead{HMXB} &
  \colhead{Total}
}
\startdata
Number & 267 & 16\tablenotemark{b} & 10\tablenotemark{b} & 15 & 304 \\
$\sigma_{\perp}/\kms$\tablenotemark{c} & 43 & 42 & 45 & 23 & 42 \\
$\sigma_{\alpha}/\kms$ & 56 & 52 & 47 & 32 & 55 \\
$\sigma_{\delta}/\kms$ & 37 & 33 & 42 & 26 & 37 \\
median($v_{\perp}/\kms$) & 44 & 39 & 38 & 38 & 43 \\
mean($v_\alpha/\kms$) & --3 & 16 & 16 & 1 & --1 \\
mean($v_\delta/\kms$) & 7 & 8 & --25 & 9 & 6 \\
\hline
$\sigma_{\rm loc,\perp}/\kms$\tablenotemark{c} & 42 & 51 & 34 & 21 & 41 \\
$\sigma_{\rm loc,\alpha}/\kms$ & 53 & 50 & 42 & 31 & 52 \\
$\sigma_{\rm loc,\delta}/\kms$ & 35 & 33 & 37 & 23 & 35 \\
median($v_{\rm loc,\perp}/\kms$) & 39 & 29 & 50 & 31 & 39 \\
mean($v_{\rm loc,\alpha}/\kms$) & 10 & 11 & 33 & 17 & 10 \\
mean($v_{\rm loc,\delta}/\kms$) & 3 & 2 & --29 & --4 & 2 \\
\enddata
\tablenotetext{a}{Includes unidentified binaries.}
\tablenotetext{b}{There are 4 stars identified as both EB and SB2.}
\tablenotetext{c}{Values for $\sigma_\perp$ and $\sigma_{\rm
    loc,\perp}$ are standard deviations from the median.}
\end{deluxetable*}

We caution that K-S tests show that the difference between the 
$\vtot$ distributions of the binary populations is not statistically
significant.  However, our sample likely contains a
substantial, perhaps even dominant, contribution from non-runaway, field
stars that formed {\it in situ} \citep{Lamb2016,Oey2013}, which
will significantly dilute the non-compact binary population in our
sample at the lowest velocities.  The fact that our HMXBs have smaller
velocity dispersions than non-compact binaries is consistent with the
expectation that bound compact binaries represent systems with less
energetic SN kicks that failed to unbind the components.  Moreover,
dynamical ejections from dense clusters can accelerate runaways to
higher velocities than the SN mechanism, since cluster acceleration
can leverage the gravitational energy from multiple stars.
Models by, e.g., \citet{Brandt1995}
and \citet{Renzo2018}
show that HMXBs have runaway
velocities $< 100\ \kms$, and typically half that value, depending on
the assumed kick velocities and pre-SN orbital parameters.  In
contrast, \citet{perets2012} find that dynamically ejected runaways
from clusters having masses on the order of $10^4\ \msol$ can reach
$200\ \kms$, including significant fractions of binaries.

Despite contamination from non-runaways systems, the non-compact
binaries show velocity distributions 
that are not only larger than for the HMXBs, but also similar 
to that for unclassified field OB stars (Table~\ref{t_binaries}).
Since the latter include single-star runaways from both mechanisms,
this suggests that dynamically ejected
objects dominate over {\it in situ} field stars in the SMC.  Furthermore,
\citet{Renzo2018} predict that $\sim 14$\% of post-SN binaries
fail to disrupt, of which some fraction are observed as HMXBs.
They also expect $\sim 3$\% of post-SN binaries to generate single
runaways faster than 30 $\kms$.  These estimates have large
uncertainties, so we might expect roughly similar
numbers of these two groups.  However, there are only 15 HMXBs, whereas
roughly half (134) of unclassified stars are fast runaways
(median $\vtot = 44\ \kms$).  While these numbers are subject to
various biases, the large disparity does suggest that
dynamical ejections likely dominate. 
We will examine additional properties, including frequencies, masses,
and rotation of these field OB runaways in future work.

\acknowledgements

We thank Roeland van der Marel for the use of his geometric correction
code, and Michal K. Szymanski for help using OGLE data in preparatory
work for this study.  MSO appreciates helpful discussions and generous
hospitality from Kaitlin Kratter and the University of Arizona in
hosting an extended visit supporting this collaboration.
We thank the anonymous referee for helpful comments and suggestions.
This work was funded by the National Science Foundation, grant AST-1514838 to MSO and
the University of Michigan.  MM acknowledges support from
NASA's Einstein Postdoctoral Fellowship program PF5-160139 and NASA
ATP grant 17-ATP17-0070; NK is supported by NSF CAREER award AST-1455260.  

\vspace{5mm}
\facilities{GAIA, Magellan, astropy}

\end{document}